\documentclass[twocolumn,prd,amsmath,amssymb]{revtex4}
\usepackage{graphicx}
\usepackage{dcolumn}
\usepackage{bm}
\usepackage{color}
\begin{document}
\title{Spin asymmetry for proton-deuteron Drell-Yan process with tensor-polarized deuteron}
\author{S. Kumano$^{1,2,3}$ and Qin-Tao Song$^{1,3}$}
\email{qintao@post.kek.jp}
\affiliation{
\baselineskip 2 pt
\hspace{1mm}
\centerline{{ $^1$ KEK Theory Center, Institute of Particle and Nuclear Studies, 
                   KEK 1-1, Oho, Tsukuba, Ibaraki, 305-0801, Japan}}\\
\centerline{{ $^2$ Particle and Nuclear Physics Division, J-PARC Center, 
                   203-1, Shirakata, Tokai, Ibaraki, 319-1106, Japan}} \\
\centerline{{ $^3$ Department of Particle and Nuclear Physics, 
                   Graduate University for Advanced Studies (SOKENDAI),}}  \\
\centerline{{1-1, Oho, Tsukuba, Ibaraki, 305-0801, Japan}}  }
% \date{\today}
\date{January 31, 2017}

\begin{abstract}
Tensor structure of the deuteron can be studied by deep inelastic scattering 
and Drell-Yan process to understand it in terms of quark and gluon degrees of freedom.
It probes interesting features in the deuteron including a D-wave contribution.
In the charged-lepton DIS, twist-two structure functions $b_1$ and $b_2$ are expressed
by tensor-polarized parton distribution functions (PDFs). 
We note that the HERMES experimental measurement of $b_1$ seems 
to be much different from a conventional theoretical prediction.
This puzzling situation should be significantly improved in the near future 
by an approved experiment to measure $b_1$ at JLab.
On the other hand, the tensor-polarized PDFs, especially antiquark distributions, 
could be measured by spin asymmetries in the Drell-Yan process with a tensor-polarized
deuteron at Fermilab. In this work, we estimate tensor-polarization
asymmetries for the Fermilab Drell-Yan experiment by using a parametrization for
the tensor-polarized PDFs to explain the HERMES $b_1$ data. Obtained spin asymmetries
are typically a few percent order and they could be measured by the Fermilab-E1039 experiment.
Since the tensor-polarized antiquark distributions will play an important role
to solve the puzzle, further theoretical and experimental efforts are needed 
toward the Drell-Yan experiment at Fermilab and other hadron facilities.
\end{abstract}
\maketitle
%%%%%%%%%%%%%%%%%%%%%%%%%%%%%%%%%%%%%%%%%%%%%%%%%%%%%%%%%%%%%%%%%%%%%%%%%%%%%%%%
\section{Introduction}

Deuteron structure has been studied by hadron degrees of freedom.
The deuteron is a bound state of proton and neutron mainly in S wave. 
In fact, the experimental magnetic moment of deuteron supports the S-wave idea,
whereas the existence of a finite electric quadrupole moment indicates 
that the deuteron should also contain D wave. Therefore, the deuteron is
an S-D mixture state, and the D-wave contribution is very small so as to
be consistent with the magnetic and quadrupole moments.

It is interesting to investigate the tensor structure in terms of 
quark and gluon degrees of freedom.
About 10 years ago,  the HERMES collaboration made the first measurement 
of the tensor structure function $b_1$ for the deuteron \cite{Airapetian:2005cb}. 
However, the measurement shows that $b_1$ is much larger than the convolution-model
prediction with the S-D mixture 
\cite{Hoodbhoy:1988am, Jaffe:1988up, Khan:1991qk, Cosyn2017}.
It indicates that the tensor structure of deuteron is not understood
in the parton level. There are other theoretical works on the deuteron $b_1$
by including shadowing phenomena, pions, and hidden-color state
\cite{Nikolaev:1996jy, Edelmann:1997qe, Bora:1997pi, Miller:2013hla}. 

There is an approved experiment to measure $b_1$ by the electron deep inelastic 
scattering (DIS) at JLab (Thomas Jefferson National Accelerator Facility) 
and it will start in a few years. 
This accurate experiment will help us to understand the tensor 
structure of deuteron. The structure function $b_1$ is expressed by the tensor-polarized 
parton distribution functions (PDFs); however, the separation of antiquark distributions 
is not obvious solely from the DIS measurements.
Since the understanding of the tensor-polarized antiquark distributions 
could be essential for clarifying the discrepancy between the conventional theory 
and the HERMES data, it is important to measure them experimentally.
Fortunately, it is possible in the Fermilab-E1039 experiment by the Drell-Yan process 
with a tensor-polarized deuteron target. The purpose of our research is to calculate 
the tensor-polarized spin asymmetries of the Drell-Yan process \cite{Kumano:2016ude} 
because there was no theoretical estimate to be used for an experimental 
proposal and future comparison with the data.

%%%%%%%%%%%%%%%%%%%%%%%%%%%%%%%%%%%%%%%%%%%%%%%%%%%%%%%%%%%%%%%%%%%%%%%%%%%%%%%%
\section{Tensor structure functions in DIS with polarized deuteron}

%%%%%%%%%%%%%%%%%%%%%%%%%%%%%%
\begin{figure}[b]
\centering
 \includegraphics[width=0.38\textwidth]{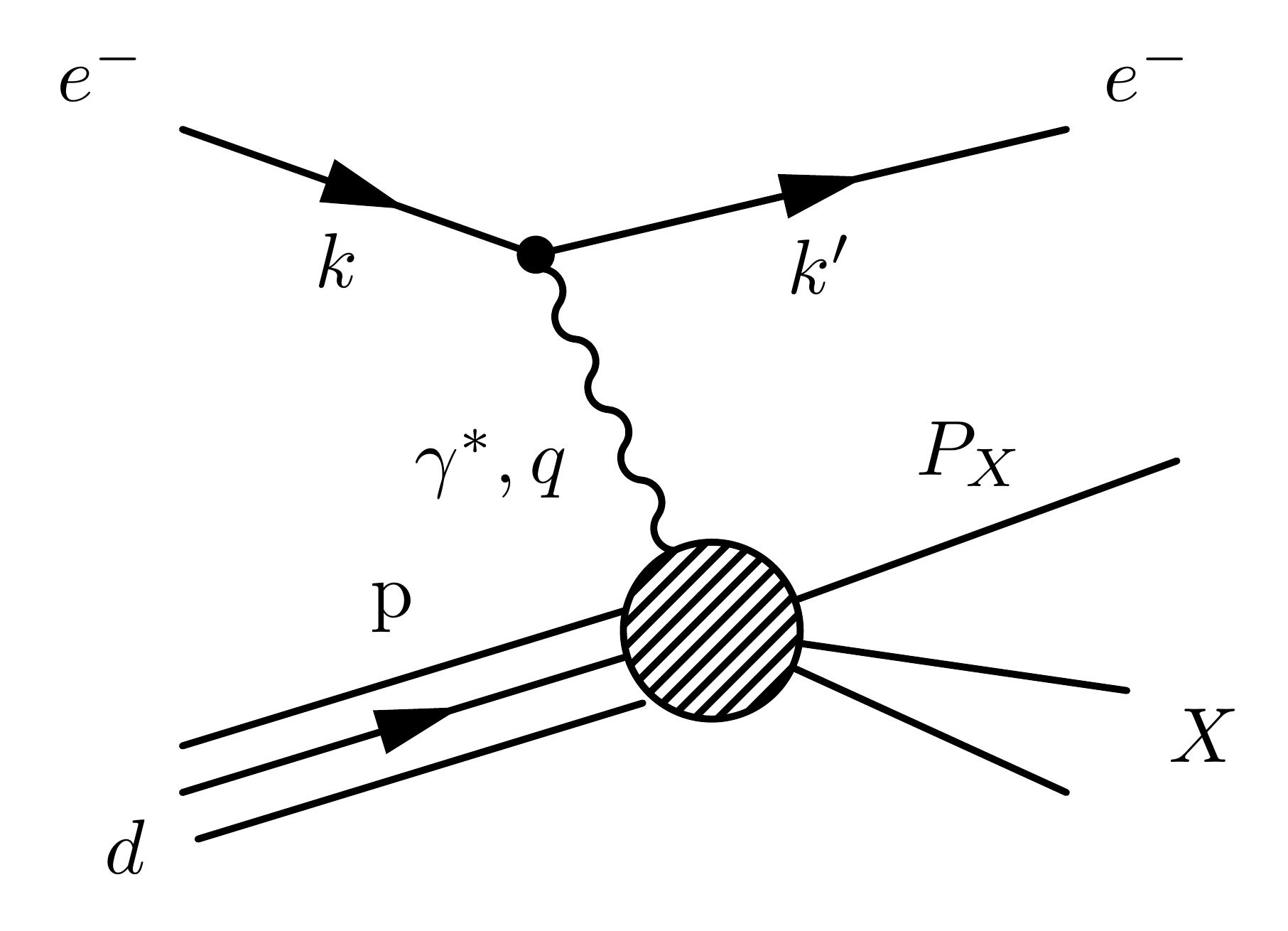}
\caption{Deep inelastic scattering with polarized deuteron.}
\label{dis}
\end{figure}
%%%%%%%%%%%%%%%%%%%%%%%%%%%%%%

The tensor structure of the deuteron can be investigated in charged-lepton DIS 
with the polarized deuteron, and it is shown in the Fig. \ref{dis}.
The hadron tensor of the deuteron is defined as
\begin{align}
W_{\mu \nu}^{\lambda_f \lambda_i}
=& \int \frac{d^4x}{4 \pi M}  e^{iqx}   \left \langle p \, \lambda_f 
       |J_{\mu}(x) J_{\nu}(0)   | p \, \lambda_i  \right \rangle   \notag \\
= & -F_1 \hat{g}_{\mu \nu} 
     +\frac{F_2}{M \nu} \hat{p}_\mu \hat{p}_\nu 
     + \frac{ig_1}{\nu}\epsilon_{\mu \nu \lambda \sigma} q^\lambda s^\sigma  \notag \\
     &     +\frac{i g_2}{M \nu ^2}\epsilon_{\mu \nu \lambda \sigma} 
      q^\lambda (p \cdot q s^\sigma - s \cdot q p^\sigma )
\notag \\
& 
     -b_1 r_{\mu \nu} 
     + \frac{1}{6} b_2 (s_{\mu \nu} +t_{\mu \nu} +u_{\mu \nu}) \notag \\
     &     + \frac{1}{2} b_3 (s_{\mu \nu} -u_{\mu \nu}) 
     + \frac{1}{2} b_4 (s_{\mu \nu} -t_{\mu \nu}) .
\label{eqn:e1}
\end{align}
where $M$, $p$, and $q$ are deuteron mass, deuteron momentum, 
and virtual-photon momentum, $\lambda_i$ and $\lambda_f$ indicate
spin states of the deuteron, and the details of other notations 
are found in Refs. \cite{Hoodbhoy:1988am,Kumano:2014pra}.

There are eight structure functions in Eq.\,(\ref{eqn:e1}).
The structure functions $F_1$, $F_2$, $g_1$ and $g_2$ exist in the spin-1/2 nucleon, 
whereas $b_1$, $b_2$, $b_3$ and $b_4$ are new structure functions 
for the spin-1 deuteron.
In the patron picture, $b_1$ is expressed by the tensor-polarized PDFs 
$\delta_Tq_i =q^0_i-(q_i^{+1}+q_i^{-1})/2$, where the superscript ($\pm 1, 0$)
indicates the deuteron spin state and the subscript $i$ is the quark flavor,
in the similar way with $F_1$:
\begin{align}
&F_1= \frac{1}{2 }\sum_i e_i^2 \left [q_i(x,Q^2)+\bar q_i(x,Q^2)   \right ], \notag \\
&b_1= \frac{1}{2 }\sum_i e_i^2 \left [ \delta_Tq_i(x,Q^2)+\delta_T \bar q_i(x,Q^2) \right ] .
\label{eqn:e7}
\end{align}
If $b_1$ is integrated over $x$, it leads to an interesting sum rule for $b_1$.
Since the tensor-polarized valence-quark distributions do not contribute to this sum,
a finite sum should come from the tensor-polarized antiquark distributions
\cite{Close:1990zw}:
\begin{align}
\! \! \! 
\int dx b_1(x)=\frac{1}{9}\int dx \left[  4  \delta_T \bar u(x)
               +  4  \delta_T \bar d(x)+  \delta_T \bar s(x)   \right ].
\label{eqn:e2}
\end{align}
Therefore, a nonzero measurement of $b_1$ integral indicates
the existence of finite tensor-polarized antiquark distributions,
in the similar way that the Gottfried sum-rule violation indicated
a finite $\bar u - \bar d$ distribution \cite{flavor3}.
This was suggested by the HERMES collaboration \cite{Airapetian:2005cb}:
\begin{align}
\int _{0.002}^{0.85} dx b_1(x) &= \left[1.05\pm 0.34\pm 0.35 \right ]\times10^{-2}, 
\notag \\
\int _{0.002}^{0.85} dx b_1(x) &= \left[0.35\pm 0.10\pm 0.18  \right ]\times10^{-2},
\label{eqn:e3}
\end{align}
where the first integral is obtained in the measured kinematical range
and the second one by imposing the constraint $Q^2 >1$ GeV$^2$.
It should be also noted that the measured HERMES $b_1$ values are
much larger in magnitude than the standard convolution-model estimates,
and it may be considered as a deuteron tensor-structure puzzle.
In any case, the antiquark distributions can be measured directly
in the Drell-Yan process, which could lead clarification of 
the puzzle.

%%%%%%%%%%%%%%%%%%%%%%%%%%%%%%%%%%%%%%%%%%%%%%%%%%%%%%%%%%%%%%%%%%%%%%%%%%%%%%%%
\section{Spin asymmetry in the Drell-Yan process with unpolarized proton 
and tensor-polarized deuteron}

%%%%%%%%%%%%%%%%%%%%%%%%%%%%%%
\begin{figure}[b]
\centering
 \includegraphics[width=0.43\textwidth]{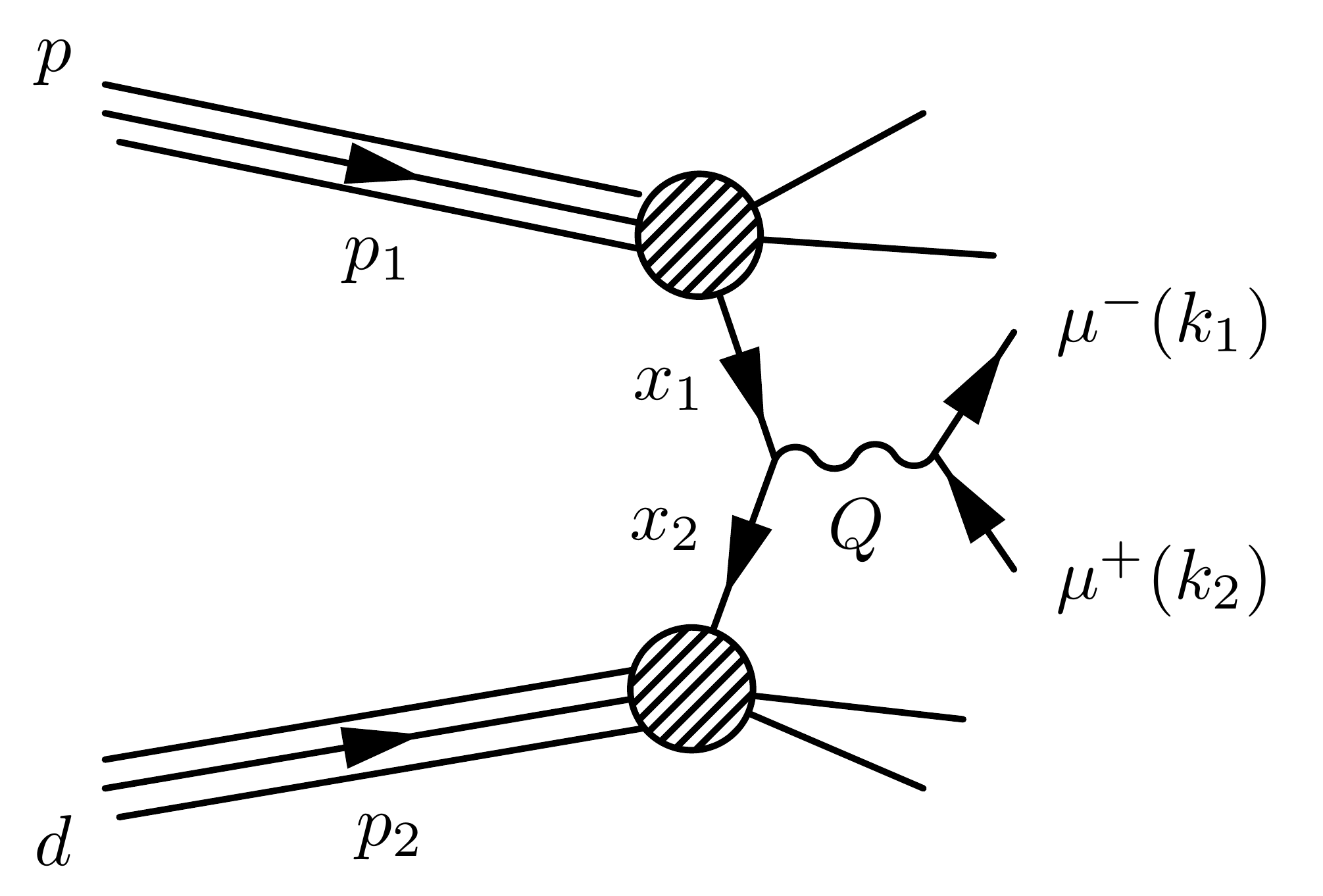}
\vspace{-0.1cm}
\caption{Drell-Yan process with unpolarized proton 
and tensor-polarized deuteron.\label{dy}}
\end{figure}
%%%%%%%%%%%%%%%%%%%%%%%%%%%%%%

The proton-deuteron Drell-Yan process is schematically shown in Fig.\,\ref{dy}, 
and the hadron tensor is defined as
\begin{align}
W_{\mu\nu}  \! =  \! \frac{1}{4\pi M} \! 
   \int  \! d^{\,4} \xi \, e^{-iQ\cdot \xi}
   \langle \, p \, d \, | \, J_\mu^{\,em} (\xi) J_\nu^{\,em}(0) \, 
   | \, p \, d \, \rangle  .
\label{eqn:dy}
\end{align}
This hadron tensor is much complicated in comparison with that of DIS, 
because there exists more than 100 structure functions 
in the polarized Drell-Yan processes. Among spin asymmetries,
$A_{UQ_0}$ \cite{ Hino:1998ww, Hino:1999qi} is the most important 
asymmetry for probing the deuteron tensor structure, and it is expressed as
\begin{align}
A_{UQ_0}=\frac{1}{2\left \langle \sigma  \right \rangle} 
\left [ \sigma(\bullet, 0)-\frac{\sigma(\bullet, +1)+\sigma(\bullet, -1)}{2}  \right],
\label{eqn:e4}
\end{align}
where $\pm$ and $0$ are the spin states of the deuteron, and $\bullet$ 
indicates the unpolarized proton. Namely, the spin asymmetry $A_{UQ_0}$
shows the cross section difference with different deuteron spin states, 
and it will disappear if the deuteron were in purely S wave.
In the following discussions, we show the asymmetry multiplied by the factor of 2 
($A_{Q} \equiv 2 A_{UQ_0}$).

In the parton model, $A_{Q}$ is related with the tensor-polarized PDFs 
of the deuteron as
\begin{align}
A_{Q}=\frac{\sum_i e_i^2 \left [q_i(x_1) \delta_T \bar q_i(x_2)
+\bar q_i(x_1) \delta_T q_i(x_2) \right]}{\sum_i e_i^2 
\left [q_i(x_1)\bar q_i(x_2)+\bar q_i(x_1)  q_i(x_2) \right]}.
\label{eqn:e4}
\end{align}
At large $x_F \, (=x_1-x_2)$, the terms $\bar q_i(x_1) \delta_T q_i(x_2)$ and 
$ \bar q_i(x_1) q_i(x_2)$ can be neglected in comparison with
$q_i(x_1) \delta_T \bar q_i(x_2)$ and
$q_i(x_1) \bar q_i(x_2) $, respectively,
and the asymmetry becomes simpler
\begin{align}
A_{Q}=\frac{\sum_i e_i^2 \left [q_i(x_1) \delta_T \bar q_i(x_2) \right]}
{\sum_i e_i^2 \left [q_i(x_1)\bar q_i(x_2) \right]}.
\label{eqn:e5}
\end{align}
Therefore, the tensor-polarized antiquark distributions can be obtained 
by measuring $A_{Q}$, and this is the advantage of using the Drell-Yan process.

%%%%%%%%%%%%%%%%%%%%%%%%%%%%%%%%%%%%%%%%%%%%%%%%%%%%%%%%%%%%%%%%%%%%%%%%%%%%%%%%
\section{Results}

Here, we present the theoretical estimates for the spin asymmetries $A_{Q}(x_1, x_2)$ 
\cite{Kumano:2016ude} of proton-deuteron Drell-Yan process for the Fermilab E-1309 
experiment. In Fig.\,\ref{dy}, quark and antiquark annihilate into dimuon through 
the virtual photon. The dimuon momentum is given by momentum fractions 
of quark and antiquark ($x_1$ and $x_2$) as
\begin{align}
M^2_{\mu\mu}=Q^2=(k_1+k_2)^2=x_1x_2s,
\label{eqn:mm}
\end{align}
where $s=(p_1+p_2)^2$ is the center-of-mass energy.
In the  E1309 experiment of Fermilab, the beam is 120 GeV unpolarized proton 
of the Main Injector and the target is a polarized deuteron.

In order to obtain the spin asymmetries $A_{Q}(x_1, x_2)$, 
the unpolarized PDFs are taken from the MSTW code \cite{Martin:2009iq} in the leading order
of the running-coupling constant $\alpha_s$. 
As for the initial tensor-polarized PDFs, the only available choice 
is the parameterization \cite{Kumano:2010vz} based on HERMES data.
In this parameterization, two sets are provided in order to find the impact of 
tensor-polarized antiquark distributions. 
There are no tensor-polarized antiquark distributions at the initial energy scale
$Q_0^2=2.5$ GeV$^2$ in set 1, whereas finite tensor-polarized antiquark distributions 
exist in set 2 even at the initial energy scale.
The set-2 parameterization should be more reliable in the sense that it
agrees with the HERMES measurements of $b_1$.
The tensor-polarized distributions at other energy scales can be obtained by
the DGLAP (Dokshitzer-Gribov-Lipatov-Altarelli-Parisi) evolution 
equations \cite{Hoodbhoy:1988am} in the same way with 
the unpolarized PDFs \cite{bf1}.

We also provide error-band estimates in this work.
There are 3 parameters involved in the initial tensor-polarized parton distributions 
in set 2 \cite{Kumano:2010vz}.
For a physical quantity $f(x)$, its error $\delta f(x)$ is expressed as
\begin{align}
[\delta f(x)]^2=\Delta \chi^2 \sum_{i, j} 
\left[ \frac{\partial f(x)}{\partial \xi_i } \right] _{ \hat{\xi}}  H_{ij}^{-1} 
\left[   \frac{\partial f(x)}{\partial \xi_j } \right] _{ \hat{\xi}},
\label{eqn:hessian}
\end{align}
where $H_{i j}$ is the Hessian matrix, $\xi_i$ is a parameter,  
and $\hat{\xi}$ is the minimum parameter set. 
Here, we adopt $\Delta \chi^2 =1$ in showing the error bands.
Expanding $\chi^2$ around the minimum parameter set $\hat{\xi}$, 
we can express $\Delta \chi^2$ by the Hessian matrix
\begin{align}
\Delta \chi^2=\chi^2(\hat{\xi}+\delta \hat{\xi})- \chi^2(\hat{\xi})=\sum_{i, j} H_{ij}\delta \xi_i \delta \xi_j.
\label{eqn:chi}
\end{align}

%%%%%%%%%%%%%%%%%%%%%%%%%%%%%%
\begin{figure}[htb!]
\centering
 \includegraphics[width=0.37\textwidth]{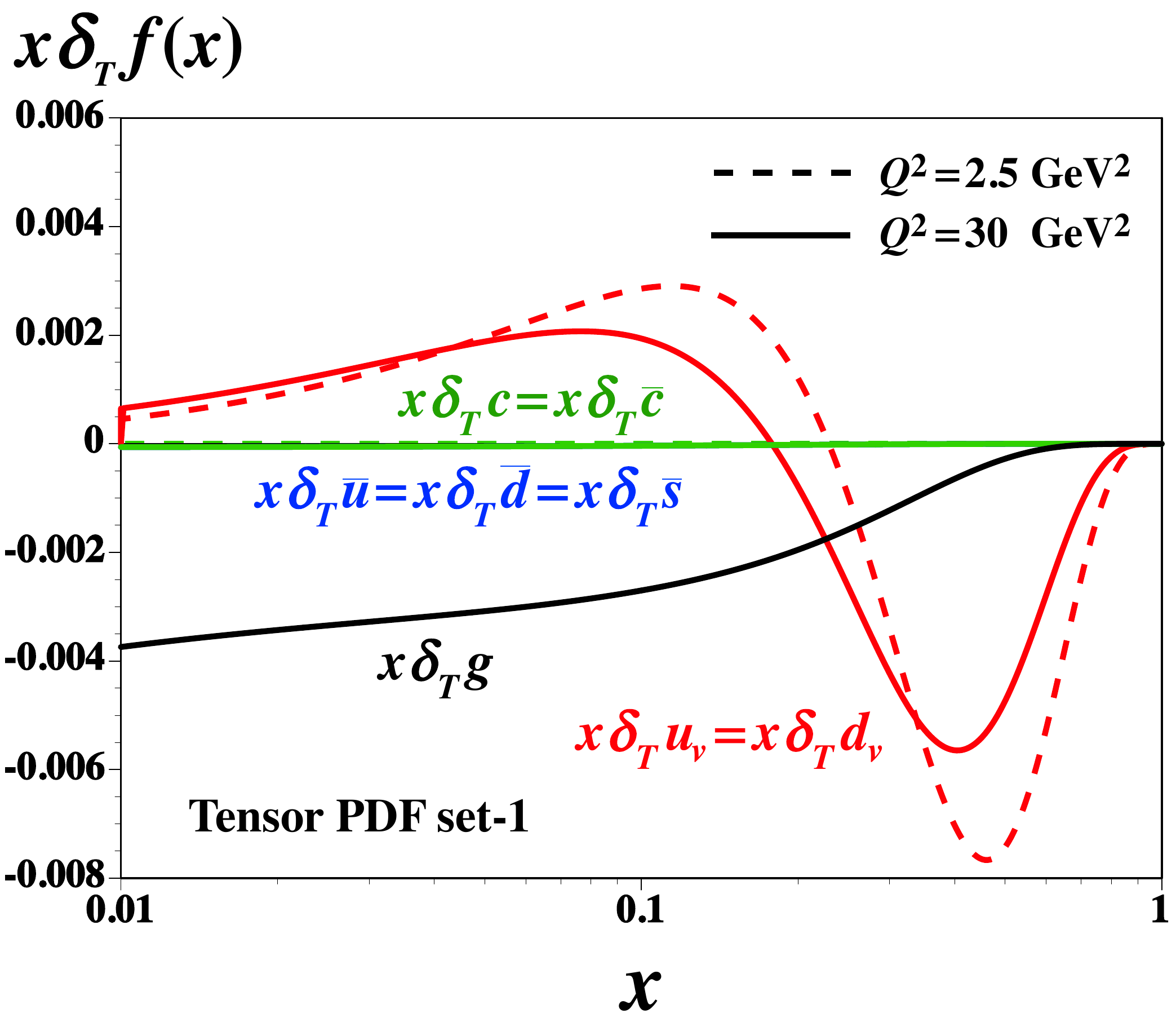}
\vspace{-0.3cm}
\caption{Tensor-polarized parton distributions 
at energy scales $Q_0^2=2.5$ GeV$^2$ (dashed curves) and $Q^2=30$ GeV$^2$ (solid curves) 
for set 1 \cite{Kumano:2016ude}.\label{tpd1}}
\end{figure}
%%%%%%%%%%%%%%%%%%%%%%%%%%%%%%
\begin{figure}[htb!]
\centering
 \includegraphics[width=0.37\textwidth]{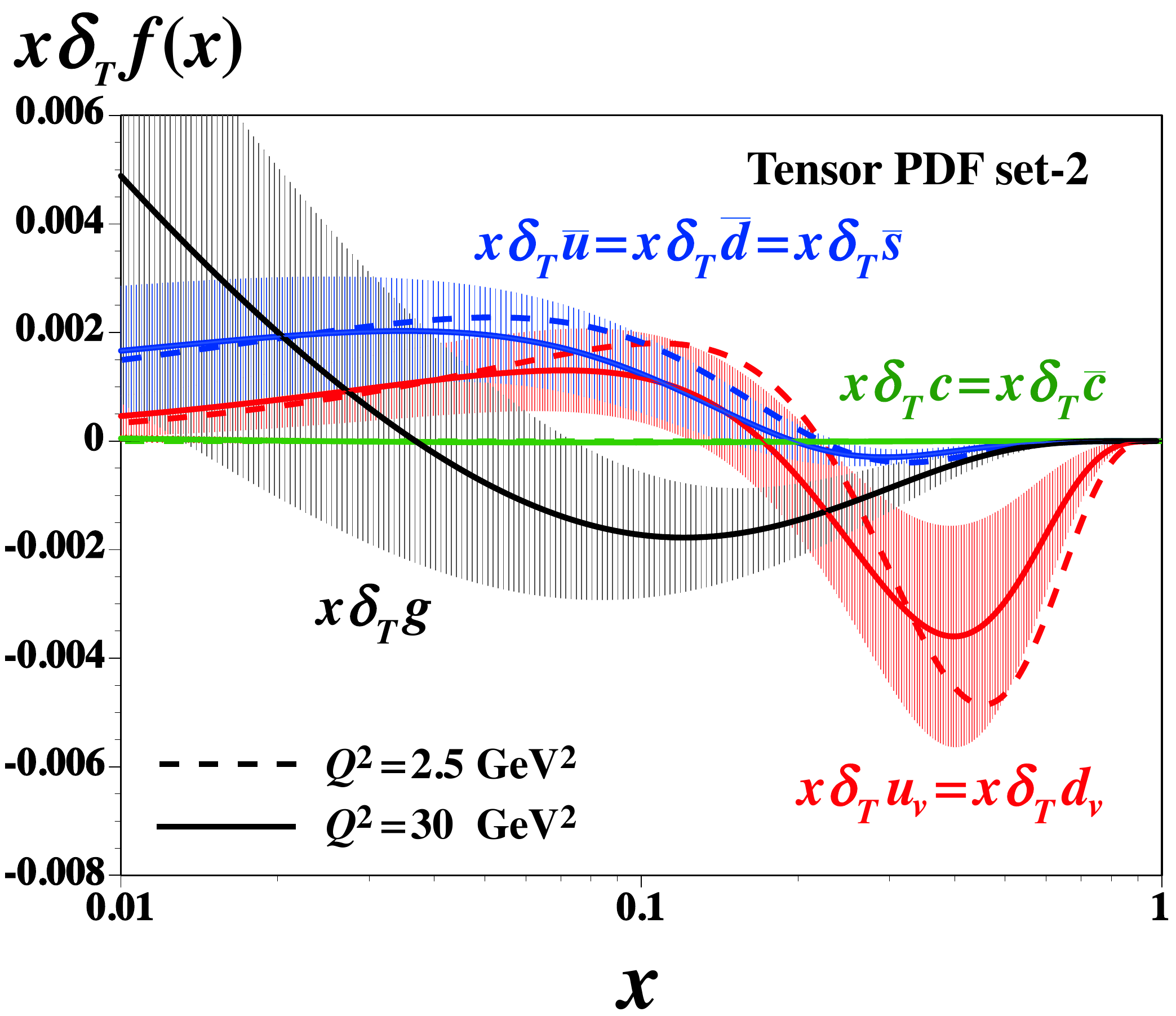}
\vspace{-0.3cm}
\caption{Tensor-polarized parton distributions with error estimates 
at energy scales $Q_0^2=2.5$ GeV$^2$ (dashed curves) and $Q^2=30$ GeV$^2$ (solid curves)  
for set 2 \cite{Kumano:2016ude}.\label{tpds}}
\vspace{-0.3cm}
\end{figure}
%%%%%%%%%%%%%%%%%%%%%%%%%%%%%%

In Fig.\,\ref{tpd1}, we show the tensor-polarized PDFs of set 1 
at the initial energy scale $Q_0^2=2.5$ GeV$^2$ and the evolved scale $Q^2=30$ GeV$^2$. 
The scale 2.5 GeV$^2$ is the average $Q^2$ value of the HERMES experiment, and
30 GeV$^2$ is roughly the average $Q^2$ value of the Fermilab Drell-Yan experiment.
Even after the $Q^2$ evolution, the tensor-polarized antiquark distributions 
are very small, since they are set to be 0 at the initial scale. 
Therefore, we have the relationship 
$\delta_T \bar u(x,Q^2)=\delta_T \bar d(x,Q^2)
=\delta_T \bar s(x,Q^2)=\delta_T \bar c(x,Q^2)$
in the $Q^2$ evolution.

The tensor-polarized PDFs of set 2 are shown in Fig.\,\ref{tpds} together 
with error bands. In Figs.\,\ref{tpd1} and \ref{tpds}, we notice that there 
also exists the tensor-polarized gluon distribution, 
even though it is set to be zero at the initial energy scale $Q_0^2=2.5$ GeV$^2$.
Because the tensor-polarized antiquark distributions are assumed 
to be equal at the initial energy scale, the relation
$\delta_T \bar u(x,Q^2)=\delta_T \bar d(x,Q^2)=\delta_T \bar s(x,Q^2)$ 
still holds in the $Q^2$ evolution.
With these $Q^2$ evolved distributions, we are now ready to calculate
tensor-polarized spin asymmetries in the Fermilab Drell-Yan experiment.

%%%%%%%%%%%%%%%%%%%%%%%%%%%%%%
\begin{figure}[t]
\centering
 \includegraphics[width=0.40\textwidth]{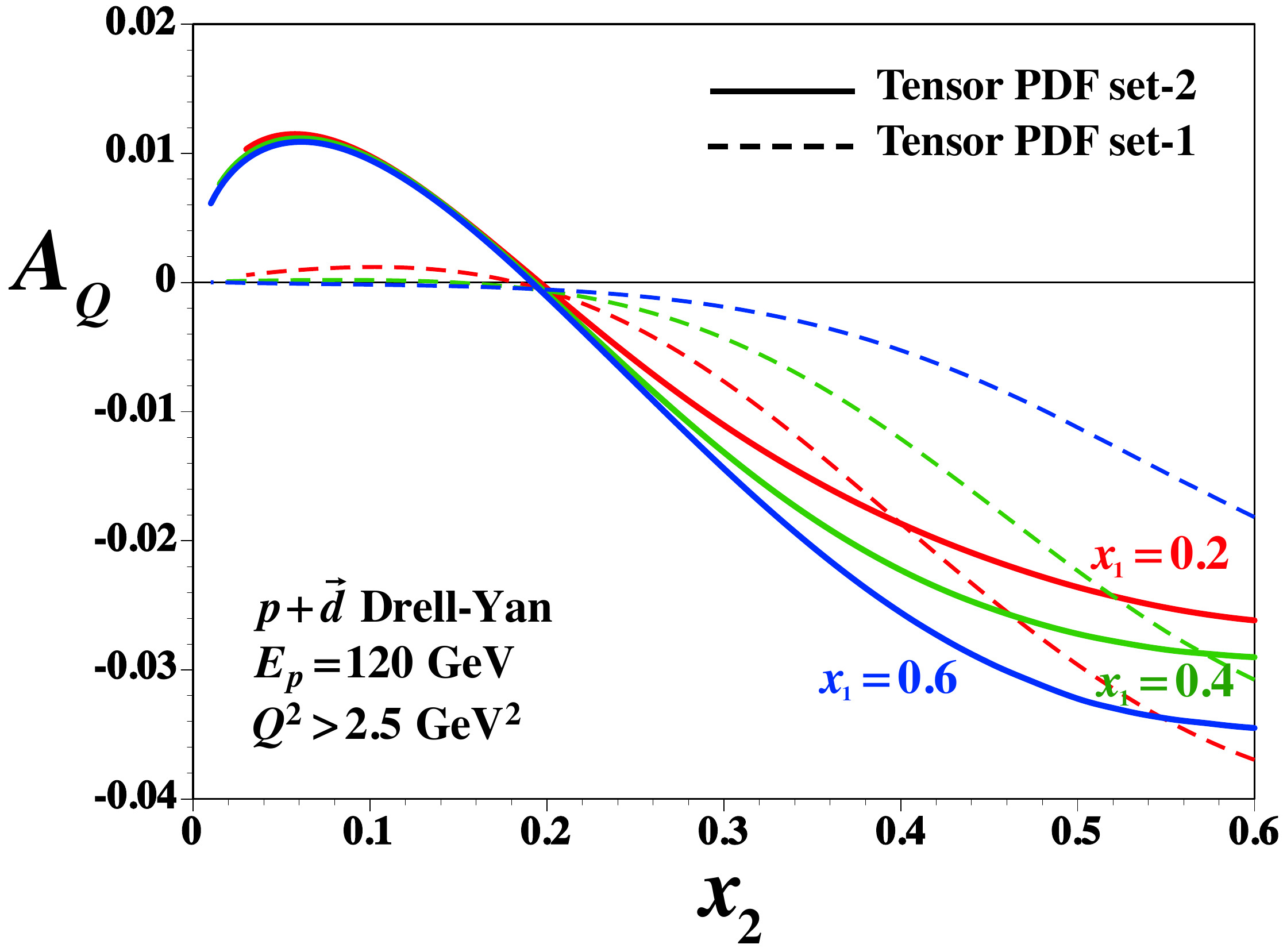}
\vspace{-0.3cm}
\caption{Spin asymmetries $A_Q(x_1, x_2)$  are estimated 
    at typical momentum fractions $x_1=0.2$, $x_1=0.4$, and $x_1=0.6$. 
    The dashed curves are for set 1 and the solid curves are 
    for set 2 \cite{Kumano:2016ude}. \label{dy-aq} }
\end{figure}
%%%%%%%%%%%%%%%%%%%%%%%%%%%%%%
\begin{figure}[htb!]
\centering
 \includegraphics[width=0.40\textwidth]{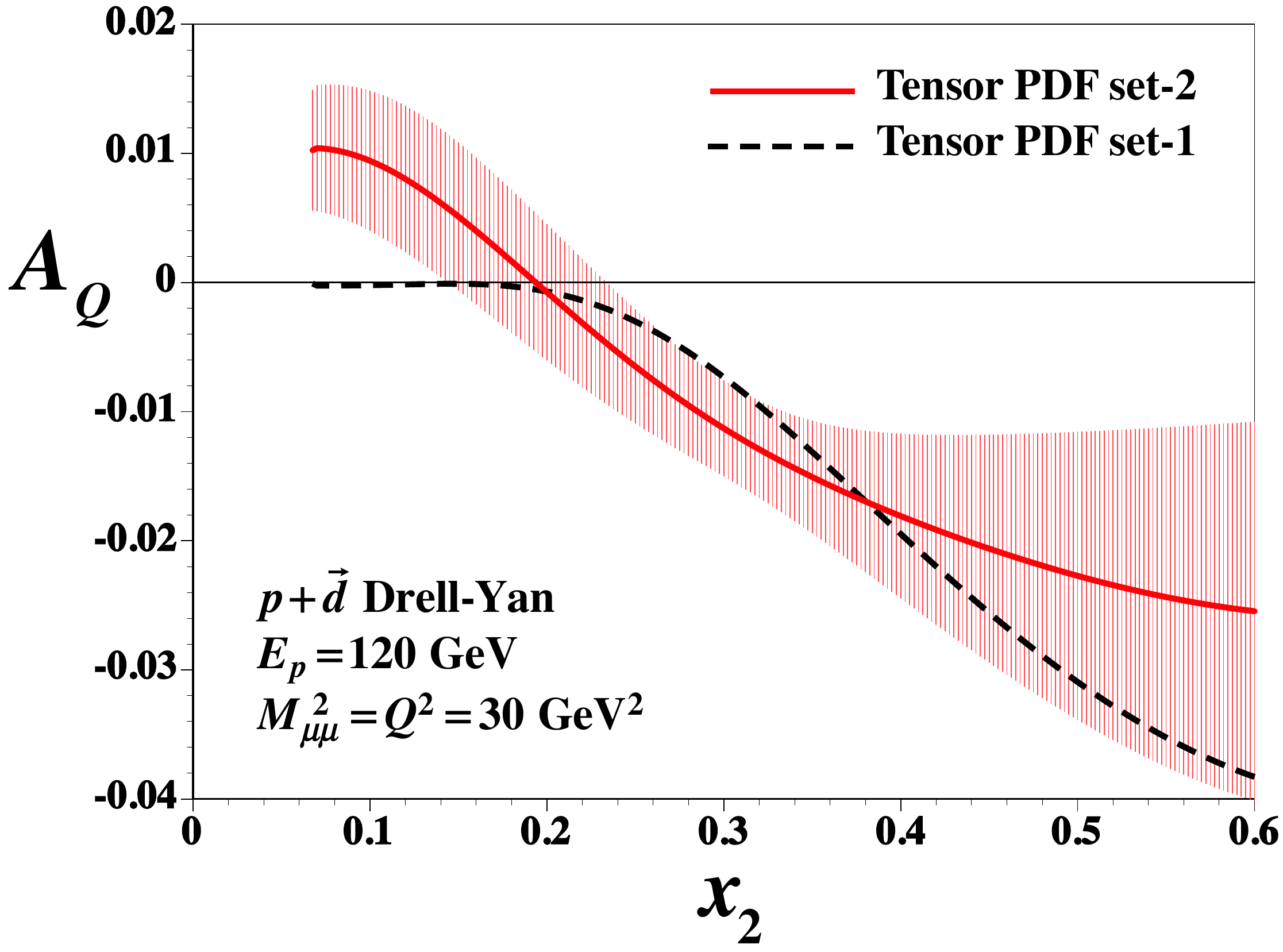}
\vspace{-0.3cm}
\caption{Spin asymmetries $A_Q(x_1, x_2)$ are estimated 
    at the typical energy scale $Q^2=30$ GeV$^2$. The dashed curve 
    is for set 1 and the solid curve is for set 2 \cite{Kumano:2016ude}. \label{30gev} }
\vspace{-0.3cm}
\end{figure}
%%%%%%%%%%%%%%%%%%%%%%%%%%%%%%

The spin asymmetries $A_Q(x_1, x_2)$ 
are shown in Fig.\,\ref{dy-aq} for both set 1 and set 2
at typical momentum fractions $x_1=0.2$, $x_1=0.4$ and $x_1=0.6$ \cite{Kumano:2016ude}.
We find that the spin asymmetries are a few percent for both sets. 
If $x_2$ is very small, the differences between the set-1 and the set-2 results 
are large. This is because the spin asymmetries $A_Q(x_1, x_2)$ are very sensitive to
the tensor-polarized antiquark distributions in this region as indicated in Eq.\,(\ref{eqn:e5}).
The set-2 asymmetries should be more reliable since the existence of finite tensor-polarized 
antiquark distributions is in agreement with the HERMES data.

In order to indicate typical errors of our estimates, we show the asymmetries 
$A_Q(x_1, x_2)$ in Fig.\,\ref{30gev} at the typical energy scale $Q^2=30$ GeV$^2$. 
Even if the error bands are considered, the asymmetries are of the order of a few percent
and they obviously deviate from 0. It validates the importance of the Fermilab Drell-Yan
experiment to probe the tensor structure of the deuteron, in particular the tensor-polarized
antiquark distributions.
In addition to the Fermilab-E1039 experiment, such a Drell-Yan experiment is possible
at hadron-accelerator facilities such as BNL-RHIC, CERN-COMPASS, J-PARC,
GSI-FAIR, and IHEP in Russia. The structure function $b_1$ can be measured
also at the future EIC. The studies of the tensor structure functions will become
one of interesting topics of hadron physics in a few years, much theoretical studies
are needed to clarify the tensor structure in the quark-gluon degrees of freedom.

$ \ \ \ $

\vspace{-0.6cm}
%%%%%%%%%%%%%%%%%%%%%%%%%%%%%%%%%%%%%%%%%%%%%%%%%%%%%%%%%%%%%%%%%%%%%%%%%%%%%%%%
\section{Summary}
\vspace{-0.3cm}

The tensor-polarized parton distributions are important physical quantities, 
and they can reflect interesting dynamical aspects of deuteron including
the D-wave contribution. 
The tensor structure of the deuteron can be studied by DIS and Drell-Yan process, 
while it is much easier to get the tensor-polarized antiquark distributions 
in the Drell-Yan process.
In this work, the tensor-polarized spin asymmetries $A_Q$ were 
theoretically calculated for the proton-deuteron Drell-Yan process
in the parton model, and we obtained a few percent values.
We also found a finite tensor-polarized gluon distribution due to $Q^2$ evolution. 
We hope that the Fermilab-E1309 experiment will be realized 
and a new field of hadron spin physics will be explored in future.

\vspace{-0.3cm}
%%%%%%%%%%%%%%%%%%%%%%%%%%%%%%%%%%%%%%%%%%%%%%%%%%%%%%%%%%%%%%%%%%%%%%%%%%%%%%%%
\begin{acknowledgements}
\vspace{-0.3cm}

This work was partially supported by JSPS KAKENHI Grant Number JP25105010. 
Q.-T. Song is supported by the MEXT Scholarship for foreign students through 
the Graduate University for Advanced Studies.
\end{acknowledgements}

%%%%%%%%%%%%%%%%%%%%%%%%%%%%%%%%%%%%%%%%%%%%%%%%%%%%%%%%%%%%%%%%%%%%%%%%%%%%%%%%

%%%%%%%%%%%%%%%%%%%%%%%%%%%%%%%%%%%%%%%%%%%%%%%%%%%%%%%%%%%%%%%%%%%%%%%%%%%%%%%%

\end{document}